\documentclass[preprint]{aastex}

\newcommand{\DD}[2]{\frac{d #1}{d #2}}
\newcommand{\DP}[2]{\frac{\partial #1}{\partial #2}}

\newcommand{\ibm}[1]{\mbox{\boldmath$#1$}}

\slugcomment{}

\shorttitle{Two-fluid Instability }
\shortauthors{Ishitsu, Inutsuka and Sekiya}

\begin{document}


\title{Two-fluid Instability of Dust and Gas in the Dust Layer of a Protoplanetary Disk}

\author{Naoki Ishitsu\altaffilmark{1}, Shu-ichiro Inutsuka\altaffilmark{2}
and Minoru Sekiya\altaffilmark{1}}
\altaffiltext{1}{Department of Earth and Planetary Sciences,
Faculty of Sciences,
33 Kyushu University, Hakozaki, Fukuoka, 812-8581, Japan }
\email{ishitsu@geo.kyushu-u.ac.jp}
\altaffiltext{2}{Department of Physics, Nagoya University,
 Furocho, Nagoya, Aichi 464-8602, Japan}


\begin{abstract}
Instabilities of the dust layer in a protoplanetary disk are investigated.
It is known that the streaming instability develops and dust density concentration occurs in a situation where the initial dust density is uniform.
This work considers the effect of initial dust density gradient vertical to the midplane.
Dust and gas are treated as different fluids.
Pressure of dust fluid is assumed to be zero.
The gas friction time is assumed to be constant.
Axisymmetric two-dimensional numerical simulation was performed using the spectral method.
We found that an instability develops with a growth rate on the order of the Keplerian angular velocity even if the gas friction time multiplied by the Keplerian angular velocity is as small as 0.001.

This instability is powered by two sources: (1) the vertical shear of the azimuthal velocity, and (2) the relative motion of dust and gas coupled with the dust density fluctuation due to advection.
This instability diffuses dust by turbulent advection and the maximum dust density decreases. 
This means that the dust concentration by the streaming instability which is seen in the case of a uniform initial dust density becomes ineffective as dust density gradient increases by the dust settling toward the midplane.
\end{abstract}


\keywords{planetary systems: protoplanetary disks---solar system: formation---hydrodynamics---instabilities
}


\section{INTRODUCTION}

The first step for planetary formation requires the formation of planetesimals larger than km-size.
However, the mechanism that planetesimals are formed from dust is scarcely understood.
The two different paths to planetesimal formation have been studied.
The one is the discontinuous formation due to the gravitational instability of the dust layer 
\citep{safronov69, goldreich73, coradini81, sekiya83, yamoto04, wakita08}.
The other one is the continuous growth to planetesimals 
due to  the sticking \citep{weiden93, cuzzi93, wurm01, kornet01}.

The gravitational instability has advantage because the problem
that meter-sized dust falls toward a central star is avoided.
The gravitational instability is theoretically more tractable 
because the formation process dose not depend on 
poorly-understood surface forces of dust material.
However, if the disk is turbulent,
the critical density for the gravitational instability
cannot be reached because the dust is stirred from the midplane.
The magneto rotational instability (MRI) is a candidate for
 turbulent sources in the stage of the  planet formation \citep{balbus91}.
The ionization degree is so low in the planet forming region 
that there is possibility that the MRI does not occur there \citep{sano00}.
It is pointed that the dust layer can become turbulent 
because of the shear instability 
even if the global turbulence such as  MRI does not occur \citep{weiden80}.
The shear instability has been extensively studied analytically 
\citep{sekiya98, sekiya00, sekiya01, ishitsu02, 
ishitsu03, michikoshi06} and numerically \citep{cuzzi93, dobro99, 
johansen06b, chiang08, barranco09}.

Recently, a different type of instability in the dust layer draws attention.
The gas supported by a negative radial pressure gradient revolves slower than the Kepler velocity.
The dust revolves against the head wind.
As a result, the dust falls toward the central star
by losing the angular momentum \citep{adachi76, weiden77}.
On the other hand, the gas moves outward in the disk
by gaining the angular momentum \citep{nakagawa86}.
\citet{youdin05} has found the streaming instability can occur 
when dust and gas have relative velocities.
Furthermore, \citet{youdin07} and \citet{johansen07a}
have detailedly performed analysis and numerical simulations 
on the streaming instability
in the situation that the unperturbed dust density is spatially homogeneous
and the gravity in the axial direction of rotation is ignored.
The simulations showed that the streaming instability 
has the action which concentrates dust.
\citet{johansen07b} has presented that Ceres-sized planetesimals are 
formed by the dust concentration due to the streaming 
instability and the self-gravity
when there are meter-sized boulders in the MRI turbulence disk.
However, the concentration by the streaming instability
is only effective for dust larger than 10cm in size.
It is not known whether dust can grow up to this size.

It has been yet to be studied what instability occurs
when the dust has distribution with a vertical gradient.
It is important to examine instability in this case 
in order to understand the growth of dust.
In this work, we perform two-fluid of gas and dust with a single size,
two-dimensional simulations.
We present that instability occurs and flow transits into turbulence
if dust with cm-sized has a graded density distribution.

In \S2, the formulation is performed under assumptions
of the fluid approximation of gas and dust.
In \S3, numerical results are presented.
In \S4, we discuss the energy sources of instability
by deriving  energy equations from linearized equations.
In  \S5, we conclude.

\section{BASIC EQUATIONS}
This section gives the basic equations.
The gas friction force is characterized by the friction time $\tau_f$, 
which is the time during which the relative velocity of a dust aggregate 
and the gas becomes $1/e$. 
The friction time depends on the radius of dust aggregate $a$,
the dust solid density $\rho_s$,  the gas density $\rho_g$, and
the thermal velocity $c_{th}$.
In the Hayashi model, the friction time is given Epstein's law 
\citep{epstein24} and Stokes' law \citep{landau87}
\begin{equation}
      \tau_f \Omega_K = \left( 1 + \frac{\pi}{8} \right) 
         \frac{a \rho_s}{\rho_g c_{th}} \Omega_K
         =1.19 \times 10^{-3} \left(\frac{\rho_s }{ 1 \mbox{gcm}^{-3} } \right)
         \left(\frac{ a }{ 1 \mbox{cm} } \right) 
         \left(\frac{ r }{1 \mbox{AU}} \right)^{1.5}  \mbox{for } a << l_g,
	\label{eqn:epstein}
 \end{equation}
\begin{equation}
       \tau_f \Omega_K = \frac{2 a^2 \rho_s}{3 l_g c_{th}} \Omega_K
     =2.77 \times 10^{-4} \left(\frac{\rho_s }{1  \mbox{gcm}^{-3} } \right)
         \left(\frac{ a }{1  \mbox{cm} } \right)^{2}
         \left(\frac{ r }{1 \mbox{AU}} \right)^{-1.25} \mbox{for } a >> l_g,
	\label{eqn:stokes}
\end{equation}
where the mean free path of the gas is given by
\begin{equation}
       l_g = \frac{1}{ \sqrt{2} \rho_g / ( \mu m_H ) \sigma_{mol} }
       =2.0 f_g^{-1}
         \left(\frac{ r }{1 \mbox{AU}} \right)^{2.75} \mbox{cm },
\end{equation}
where $\mu$ is the mean molecular weight, $m_H$ is the mass of a hydrogen atom,
and $\sigma_{mol}$ is the mean cross section of the molecules,
$f_g$ is the gas density ratio compared to the Hayashi model \citep{hayashi81}.

We assume that all dust aggregates have an identical friction time, 
and treat dust aggregates as a pressure-less fluid.
The latter assumption is good only if $\tau_f \Omega_K \ll 1$ 
\citep{garaud04}. 
However we performed numerical simulations for wide rage of the 
value $\tau_f \Omega_K$ in order to understand basic physics 
of dust-gas two fluids.

We neglect the curvature of the cylindrical coordinates $(r,\phi, z)$
and use the local Cartesian coordinate system which rotates
with the Keplerian velocity. 
Our coordinates $x$, $y$, and $z$ denote radial, azimuthal, 
and vertical directions of the disk, respectively. 
That is, $x= r-R$, $y=R[\phi-\Omega_K(R)t]$, and $z$, 
where $\Omega_K(R)$ is the Keplerian angular velocity 
at a fiducial radius $r=R$ 
and we neglect higher order terms of $x$, $y$ and $z$. 
In the following, we denote $v_K(R)$ and $\Omega_K(R)$ by $v_K$
and $\Omega_K$ for simplicity.  
The gas can be assumed to be incompressible because the dust layer treated here 
is much thinner than vertical scale height of the gas disk and, 
in addition, the flow velocity is subsonic.
The vertical components of the gravity of the central star and 
the disk self-gravity are neglected; 
this assumption is also used in the previous works 
\citep{youdin05, youdin07, johansen07a}.
We consider an unperturbed state in which the radial pressure
gradient $\partial P_0/ \partial R$ is a negative constant.

We assume axisymmetric flows, i.e. physical quantities are independent of $y$. 
Thus, we obtain the continuity equations, the momentum equations 
of gas and dust,
 \begin{equation}
	 \nabla_2 \cdot \ibm{U}_g=0,
	 \label{eqn:0ba1}
 \end{equation}
\begin{equation}
	 \DP{ \ibm{U}_g}{t} + ( \ibm{U}_g \cdot \nabla_2 ) \ibm{U}_g
      = - \frac{1}{\rho_g} \nabla_2 P  
      - \frac{1}{\rho_g} \DP{P_0}{R} \ibm{\hat{x}}
      + 2 \ibm{U}_g \times \ibm{\Omega}_K
      + 3 \Omega^2_K x \ibm{\hat{x}}
      - \frac{\rho_d}{\tau_f\rho_g}(\ibm{U}_g-\ibm{U}_d),
	\label{eqn:0ba2}
\end{equation} 
\begin{equation}
	 \DP{\rho_d}{t} + \nabla_2 \cdot (\rho_d \ibm{U}_d) 
            = \nu_D \nabla^2_2 \rho_d,
	\label{eqn:0ba3}
\end{equation} 
\begin{equation}
	 \DP{\ibm{U}_d}{t} + ( \ibm{U}_d \cdot \nabla_2 ) \ibm{U}_d
      =    2 \ibm{U}_d \times \ibm{\Omega}_K   
             + 3 \Omega^2_K x \ibm{\hat{x}}
        - \frac{1}{\tau_f}(\ibm{U}_d-\ibm{U}_g),
	\label{eqn:0ba4}
\end{equation} 
where
$\nabla_2 = (\DP{}{x}, 0, \DP{}{z})$, $\ibm{\Omega}_K = (0, 0, \Omega_K)$,
$P$ is the local gas pressure perturbation,
$\rho_d$ is the dust density defined by the total dust mass floating in a unit volume,
and 
 $\ibm{U} \equiv(U, V, W)$ is velocity of gas and  dust,
the subscripts $g$ and $d$ denote gas and dust, respectively.
In order to solve the continuity equation of dust stably,
the diffusive term is added to equation(\ref{eqn:0ba3}) artificially.
The diffusive parameter $\nu_D$ is chosen so that
the numerical stability is maintained, and also
the numerical diffusion is negligibly small.
We confirmed that dust mass was conserved in this method.

We perform velocity translation 
in the azimuthal direction following \cite{johansen06a}.
The system which balances between global pressure gradient,
centrifugal force, and gravity at $r=R$ rotates with 
a sub-Kepler velocity given by 
\begin{equation}
     V_0 = - \frac{3}{2} \Omega_K x - \eta v_K,
    \label{eqn:v0}
\end{equation}
where $\eta$ is a dimensionless parameter which expresses the effect of 
global radial pressure gradient and defined by
\begin{equation}
  \eta = - \frac{1}{2\rho_g \Omega_K^2 R} \DP{P_0}{R} 
        = 1.81 \times 10^{-3} (R/\mbox{1AU})^{1/2} .
    \label{eqn:eta}
\end{equation}
If the reaction force of dust on gas through the friction is negligibly small, 
the gas revolves at  $\eta v_K = 54$ $\mbox{m s}^{-1}$
slower than the Kepler velocity in the Hayashi model disk.
Substituting $ \ibm{U} = \ibm{u} - V_0 \ibm{\hat{y}}$ 
into equations (\ref{eqn:0ba1}) -- (\ref{eqn:0ba4}) gives
\begin{equation}
	 \nabla_2 \cdot \ibm{u}_g=0,
	 \label{eqn:ba1}
 \end{equation}
\begin{equation}
	 \DP{ \ibm{u}_g}{t} + ( \ibm{u}_g \cdot \nabla_2 ) \ibm{u}_g
      = - \frac{1}{\rho_g} \nabla_2 P + 2 \ibm{u}_g \times \ibm{\Omega}_K
       + \frac{3}{2} u_g \Omega_K \ibm{\hat{y}}
      - \frac{\rho_d}{\tau_f\rho_g}(\ibm{u}_g-\ibm{u}_d),
	\label{eqn:ba2}
\end{equation} 
\begin{equation}
	 \DP{\rho_d}{t} + \nabla_2 \cdot (\rho_d \ibm{u}_d) 
            = \nu_D \nabla^2_2 \rho_d,
	\label{eqn:ba3}
\end{equation} 
\begin{equation}
	 \DP{\ibm{u}_d}{t} + ( \ibm{u}_d \cdot \nabla_2 ) \ibm{u}_d
      =    2 \ibm{u}_d \times \ibm{\Omega}_K   
             + \frac{3}{2} u_d \Omega_K \ibm{\hat{y}}
        - \frac{1}{\tau_f}(\ibm{u}_d-\ibm{u}_g) 
        - 2 \eta v_K \Omega_K \ibm{\hat{x}}.
	\label{eqn:ba4}
\end{equation} 
Equations (\ref{eqn:ba1})--(\ref{eqn:ba4}) are solved 
with the Fourier spectral method.
Boundary conditions are periodic.
As for periodic boundary condition of the $z$ direction, 
computational box size $L_z$ is large enough for dust not 
to cross the boundary and 
for the eigenfunction obtained by the normal mode analysis to decay.
A second-order Adams-Bashforth scheme for the non-linear terms and a
Crank-Nicolson scheme for the viscous terms are employed. 
We use the phase shift method to eliminate aliasing error
\citep{canuto88}.

\subsection{INITIAL CONDITIONS}
\citet{youdin05}, \citet{youdin07} and \citet{johansen07a} investigated
 the streaming instability of the dust layer in a protoplanetary disk 
using an unstratified uniform dust density distribution as their initial conditions.  
However, a dust concentrated region in which the dust density 
has the same orders of magnitude with the gas density 
would be realized if dust settle toward the midplane,
and the dust density should decrease with the distance from the midplane.  
As a simple model of this setting, 
we employ the initial density distribution 
which has constant dust density around the midplane and sinusoidal transition zones: 
\begin{equation}
\rho_{d}(z) =
\left\{ \begin{array}{rl}
\rho_{d}(0) & \mbox{ for $|z|\le z_d-2h_d$,} \\
\rho_{d}(0)\{1-\sin [\pi (z-z_d+h_d)/2h_d]\}/2 & 
\mbox{  for $z_d-2h_d<|z|<z_d$,} \\
0 & \mbox{ for $z_d\le |z|$},
\end{array}
\right.
\label{eq:rho0}
\end{equation}
where $z_d$ the half-thickness of the dust layer,
and $h_d$  the half-thickness of the transition zones,
where the dust density varies from $\rho_{d0}(0)$ to 0 sinusoidally.
Here the half-thickness of the dust layer is given by
\begin{equation}
	z_d =\frac{\Sigma_d  }{2\rho_{d}(0)} + h_d,
\end{equation}
and the surface density of the dust is given by
\begin{equation}
\Sigma_d=\int_{-\infty}^{+\infty}\rho_d dz=
\left\{ \begin{array}{rl}
7.1f_d(r/\mbox{AU})^{-1.5} \mbox{ g cm}^2\mbox{ for } r<2.8\mbox{AU},\\
30f_d(r/\mbox{AU})^{-1.5} \mbox{ g cm}^2\mbox{ for } r>2.8\mbox{AU},
	\label{eqn:Sigma}
\end{array}
\right.
\end{equation}
where $f_d$ is a parameter 
($f_d=1$ for the Hayashi model). 
We used Hayashi's solar nebula model \citep{hayashi81, hayashi85} at 1AU as the dust surface density $\Sigma_d$.
Initial velocities of dust and gas are given quasi-stationary flow 
obtained by \citet{nakagawa86}.

\begin{equation}
 	\bar{u}_g =  \frac{2 \tau_f \Omega_K \rho_d \rho_g} 
            { ( \rho_d+\rho_g)^2 + (\tau_f \Omega_K \rho_g)^2} \eta v_K,
 	 \label{eqn:nueq1}
\end{equation}
 \begin{equation}
 	\bar{v}_g = \frac{\rho_d(\rho_d + \rho_g)} 
          { ( \rho_d+\rho_g)^2 + (\tau_f \Omega_K \rho_g )^2} \eta v_K,
	\label{eqn:nueq2}
\end{equation}
 \begin{equation}
 	\bar{u}_d = - \frac{2 \tau_f \Omega_K} 
            { ( \rho_d+\rho_g)^2 + (\tau_f \Omega_K \rho_g)^2} \eta v_K,
	\label{eqn:nueq3}
\end{equation}
 \begin{equation}
 	\bar{v}_d = \left[1 - \frac{ \rho_g (\rho_d + \rho_g)}  
                    { ( \rho_d+\rho_g)^2 + (\tau_f \Omega_K \rho_g)^2}
                    \right] \eta v_K,
	\label{eqn:nueq4}
\end{equation}
 \begin{equation}
 	\bar{w}_g = \bar{w}_d =0.
	\label{eqn:nueq5}	
\end{equation}
Note that the co-ordinate system in this paper moves with sub-Kepler
velocity $(1-\eta) v_K$; on the other hand, it moves with 
the Kepler velocity $v_K$ in \citet{nakagawa86}.

\section{NUMERICAL RESULTS}
This section displays results of the numerical simulation of dust and gas, 
two-fluid simulations.
Model parameters used in this work are listed in Table 1.

\subsection{ Constant Dust Density Distribution}
\begin{figure}
\plotone{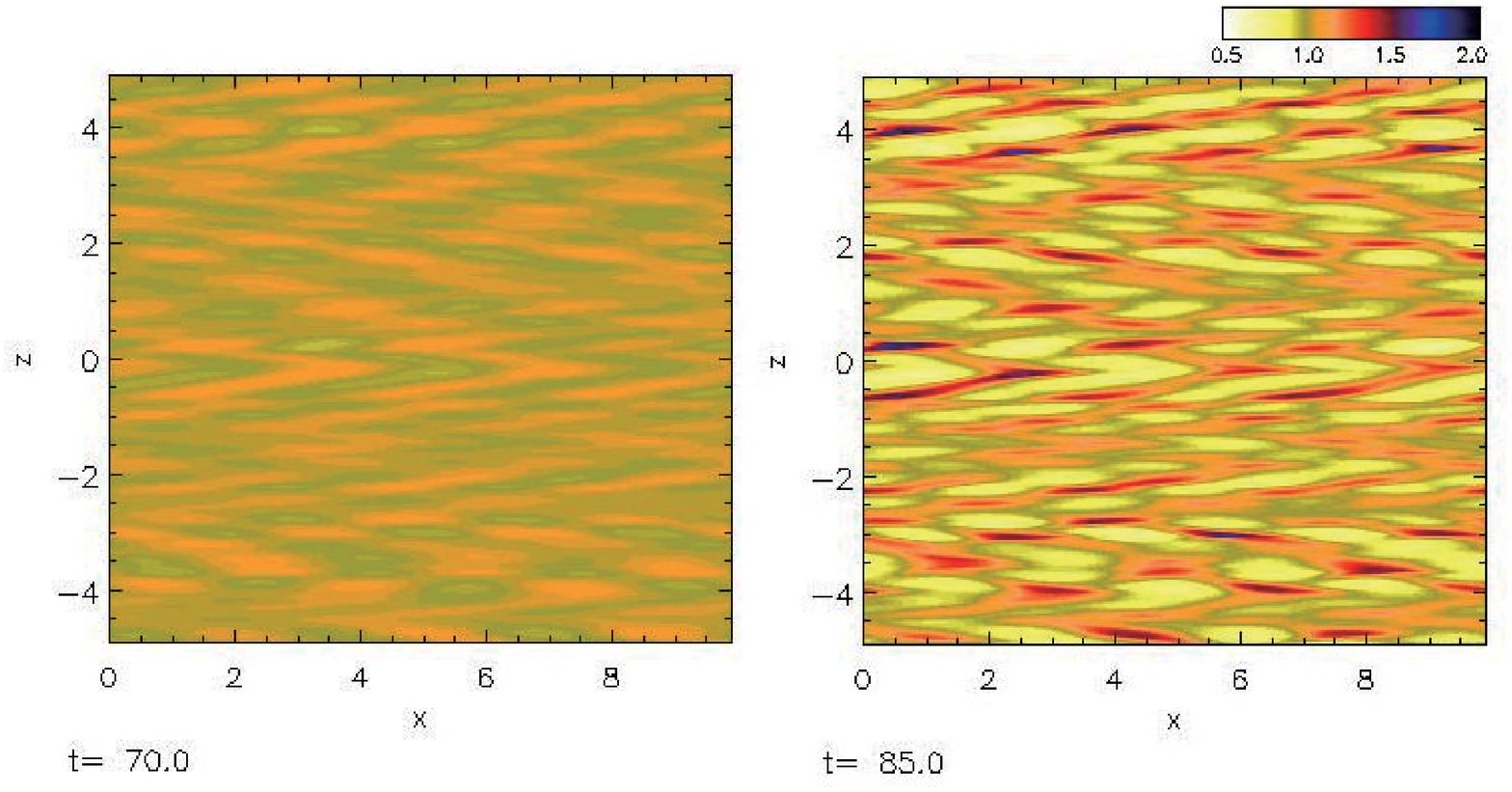}
\caption{
Snapshots of dust density at times $t \Omega_K=70$ and 85,
in the case where $\tau_f \Omega_K=1$ and  $\rho_d(0)/\rho_g =1$.
Time, length and density are normalized by $\Omega_K, \eta r$
and $\rho_g$, respectively.
\label{densityct0}}
\end{figure}

We now present a simulation result under the constant dust distribution
condition
for comparing our work to \citet{youdin07} and \citet{johansen07a}.
Figure \ref{densityct0} shows snapshots for the evolution of dust density,
where $\rho_d(0)/\rho_g=1, h_d/ z_d =0$, and $\tau_f \Omega_K=1$.
The streaming instability occurs similar to the results of \citet{johansen07a}.
The dust concentration is seen after turbulence fully develops.
Table 1 shows the growth rate $\omega_I$ and the radial wave number $k$ 
of the most unstable mode.
We performed the linear analysis based on \cite{youdin07},
and confirmed that the streaming instability is reproduced well
with an accuracy of 1 \% error in the growth rate.

\subsection{Stratified Dust Density Distribution}
\begin{figure}
\plotone{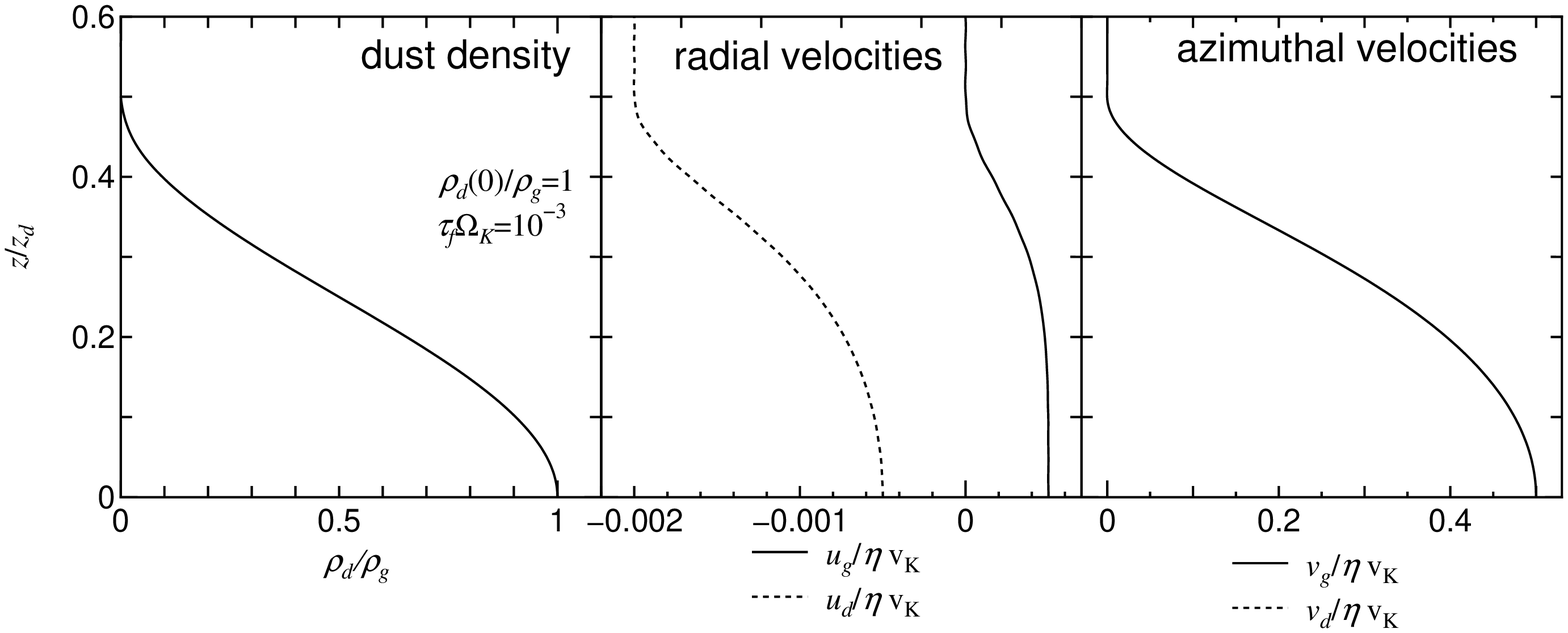}
\caption{
Distributions of dust density, gas and dust velocities of 
the initial flow
in the case where $\tau_f \Omega_K=10^{-3}$ and  $\rho_d(0)/\rho_g =1$.
In the right panel, the curve of the dust velocity is not seen
because the gas and dust have the almost same azimuthal velocity.
\label{basicr1t3}}
\end{figure}

\begin{figure}
\plotone{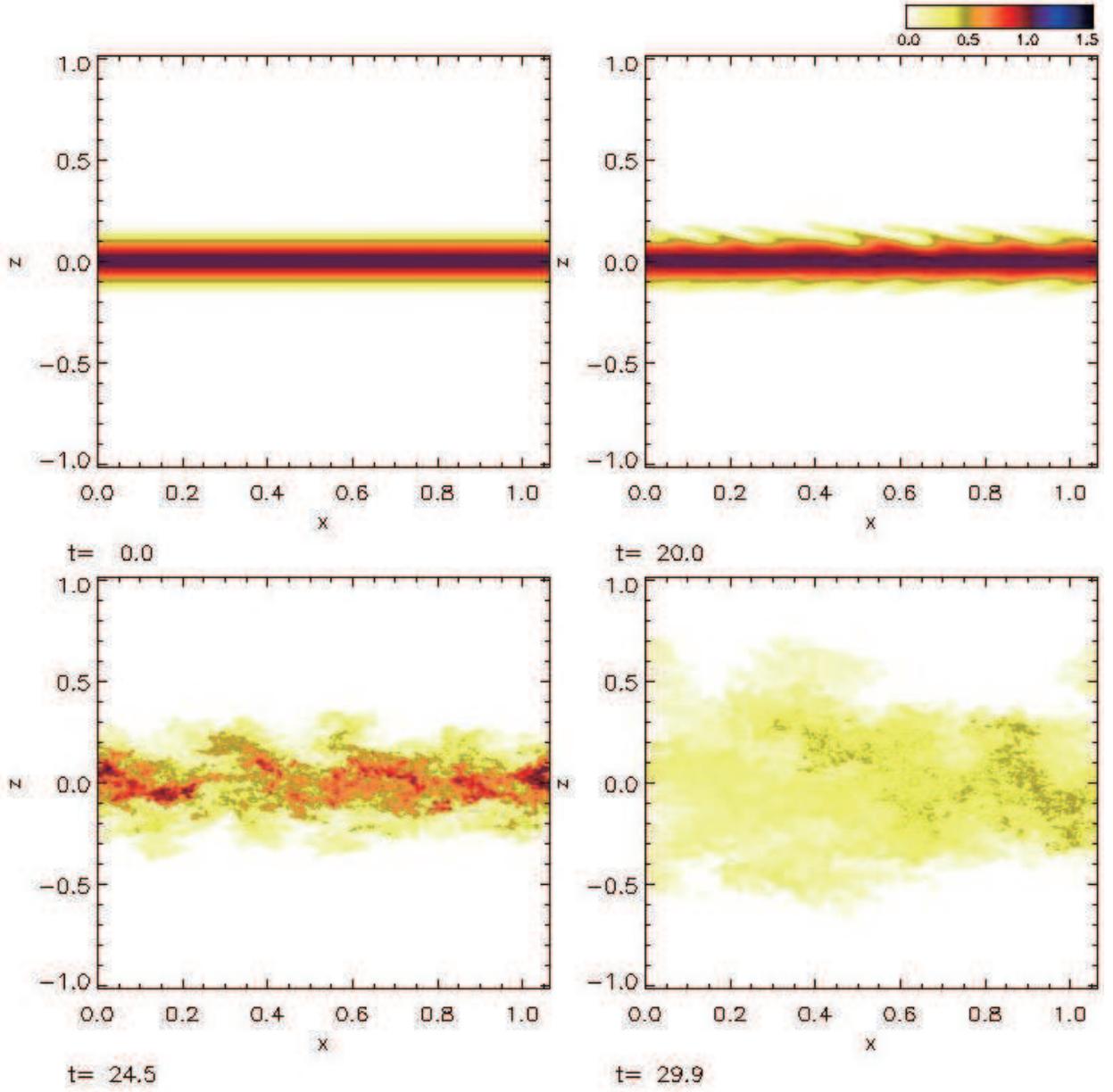}
\caption{
Snapshots of dust density at times $t \Omega_K=0, 20.0, 24.5$ and 29.9,
in the case where $\tau_f \Omega_K=10^{-3}$ and  $\rho_d(0)/\rho_g =1$.
\label{densitygt3}}
\end{figure}

Here, results for stratified dust density distributions 
are presented.
First, we investigate the case of $\tau_f \Omega_K=10^{-3}$.
This friction time corresponds to approximately 1cm dust diameter 
at 1 AU from equation (\ref{eqn:epstein}).
Figure \ref{basicr1t3} shows the initial profiles of the
dust density and velocities of gas and dust
 for $\rho_d(0)/\rho_g=1$ and $h_d/ z_d =0.5$.
It is seen that the radial velocities of dust and gas are very slow,
$ |\bar{u}_g|/(\eta v_k ), |\bar{u}_d|/(\eta v_k ) \sim 10^{-3}$. 
Figure \ref{densitygt3} shows snapshots of the evolution of dust density.
The perturbation of short wave length grows by an instability, and 
the flow transits to turbulent state.
And then, the dust is stirred from the midplane.
Because the growth time of instability $1/\omega_I \sim 1$yr
is much shorter than the settling time of dust 
$1 / (\tau_f \Omega_K^2) \sim 10^3$ yr,
the result would not change even if there is the vertical 
gravity which is omitted in our simulation.
Thus, this instability may prevent the disk from the gravitational instability.

\begin{figure}
\plotone{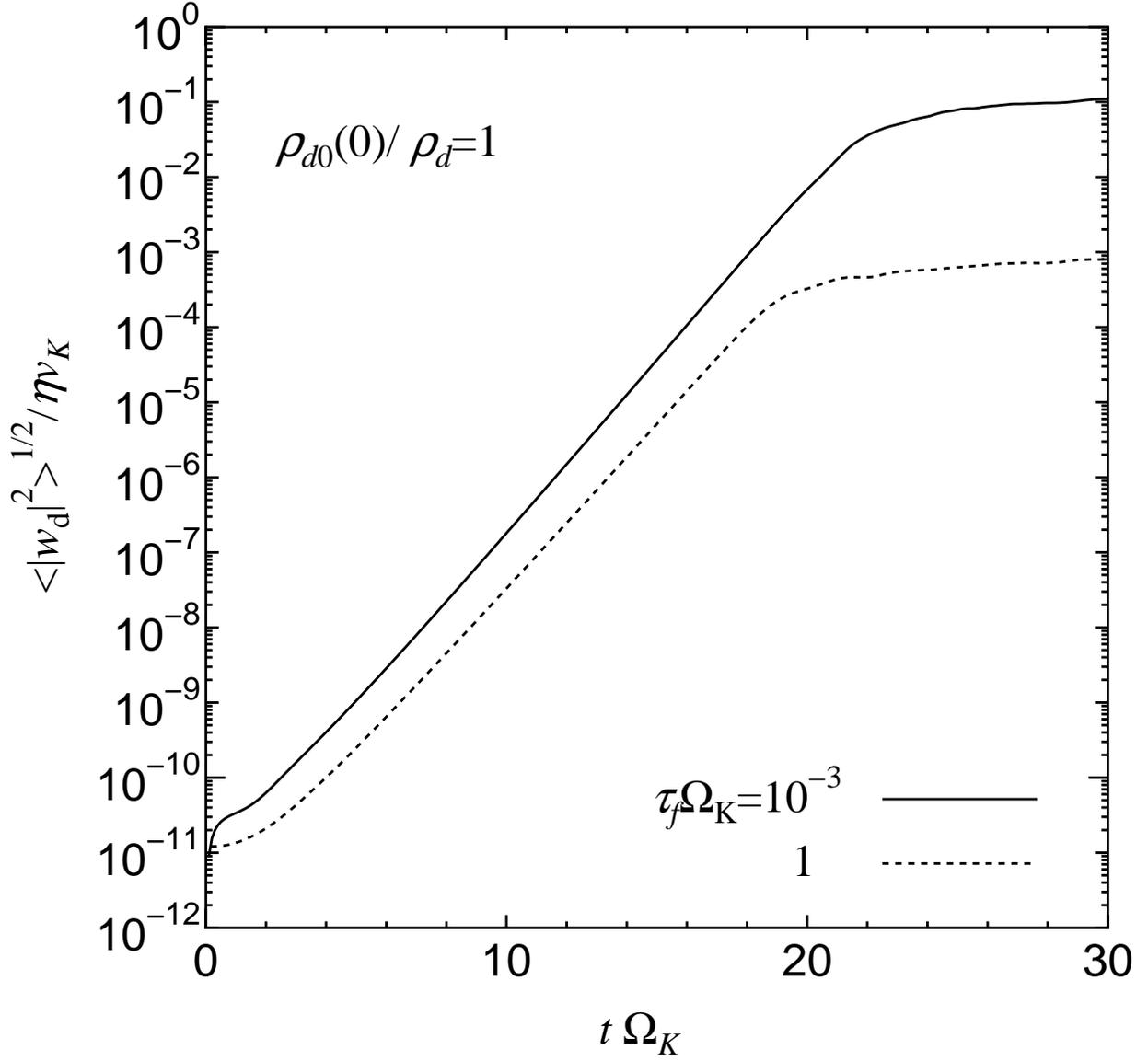}
\caption{
The time evolutions of the density weighted averag of the 
vertical dust r.m.s. defined by the equation (\ref{eqn:wdrms}),
in the case where $\tau_f \Omega_K=10^{-3}$(solid) and  1 (dotted) 
with $\rho_d(0)/\rho_g =1$.
\label{t-wdrms}}
\end{figure}

The dust concentration occurs as a result of ``streaming instability''
if the dust density distribution is constant as shown by
\citet{youdin05}, \citet{youdin07}, and \citet{johansen07a}.
However, if an initial dust density is not homogeneous,
maximum dust density does not exceed initial one.
The growth rate of instability is approximately the Keplerian frequency 
as seen in Table 1.
The detail of the instability in this case is analyzed in \S4.
Figure \ref{t-wdrms} shows the evolution of r.m.s. of
vertical dust velocities weight-averaged by dust density 
$\langle w_d \rangle_{rms}$.
Here
\begin{equation}
   \langle w_d \rangle_{rms} 
= \sqrt{\frac{\int \rho_d w_d^2 dV}{\int \rho_d dV}}.
   \label{eqn:wdrms}
\end{equation}
The r.m.s of dust velocity shows the linear-growth at first
and then saturates at $\langle w_d \rangle_{rms} \approx 0.1 \eta v_K$
(several m/sec).

\begin{figure}
\plotone{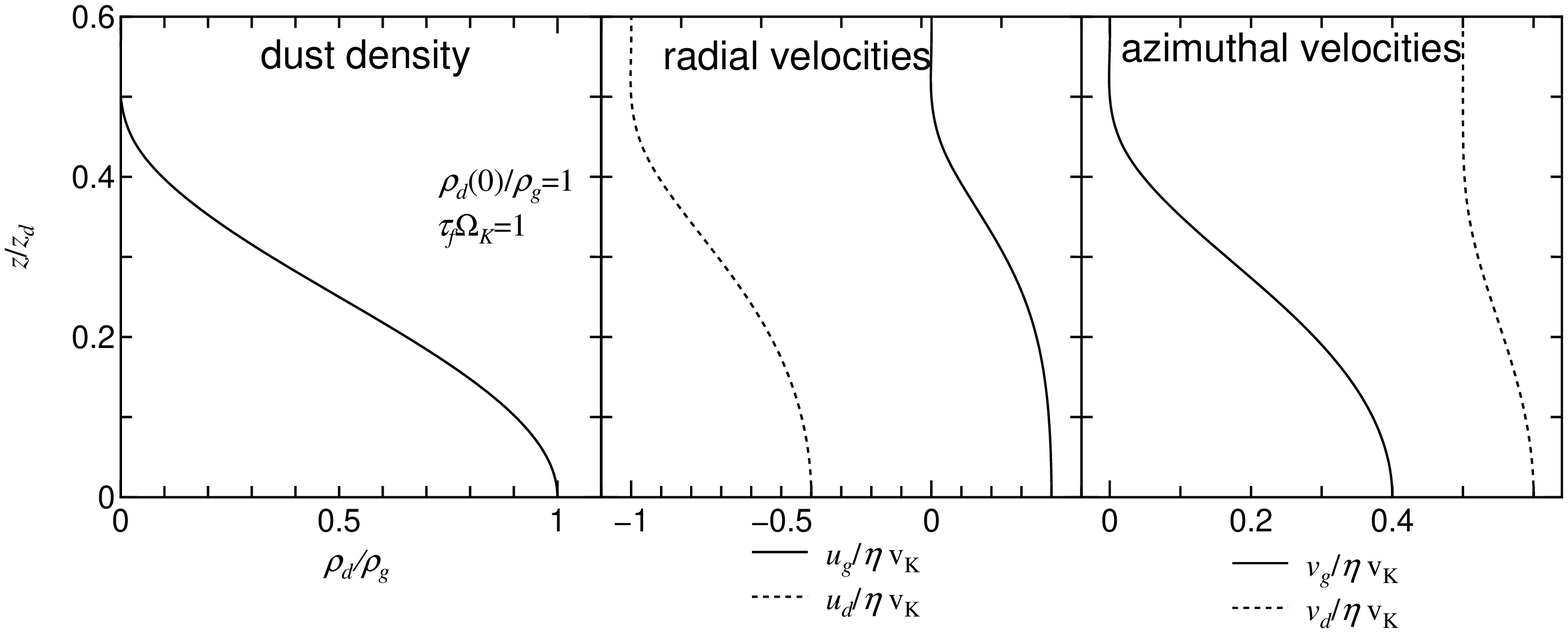}
\caption{
Distributions of dust density, gas and dust velocities of 
the initial flow
in the case where $\tau_f \Omega_K=1$ and   $\rho_d(0)/\rho_g =1$.
\label{basicr1t0}}
\end{figure}

Next we consider the case of $\tau_f \Omega_K=1$.
This friction time approximately corresponds to dust with $a=$ 1m 
at 1AU from equation (\ref{eqn:stokes}).
In this case, we expect that the random velocities of dust is large, 
so that our approximation of pressure-less fluid for dust is not justified.
However, we perform calculations for this case in order to understand 
the essence of the dust and gas two-fluid instability 
by comparing the results with that for $\tau_f \Omega_K=10^{-3}$.
Figure \ref{basicr1t0} shows the initial profiles of 
dust density and velocities of the gas and dust
for $\rho_d(0)/\rho_g=1$ and $h_d/ z_d =0.5$.
A m-sized body falls toward a central star with velocity $\eta v_K$.
As the dust settles toward the midplane to form the dust layer,
dust density grows compared with the gas density.
Then, the infall velocity of dust toward the central star 
slows down due to the increase of dust inertia as seen from equation 
(\ref{eqn:nueq3}).
On the other hand, gas moves outward
by obtaining the angular momentum from dust 
as seen from equation (\ref{eqn:nueq1}).
Radial velocities of dust and gas vary with the distance from midplane
depending on the dust density;
that is, the vertical shear of radial velocities arises.
The density pattern of the shear instability is actually seen 
at the upper right panel in Figure \ref{basicr1t0}.
It is shown in \S4
that the energy source of this  instability is vertical shear of
the azimuthal and radial components of the gas velocity.

\begin{figure}
\plotone{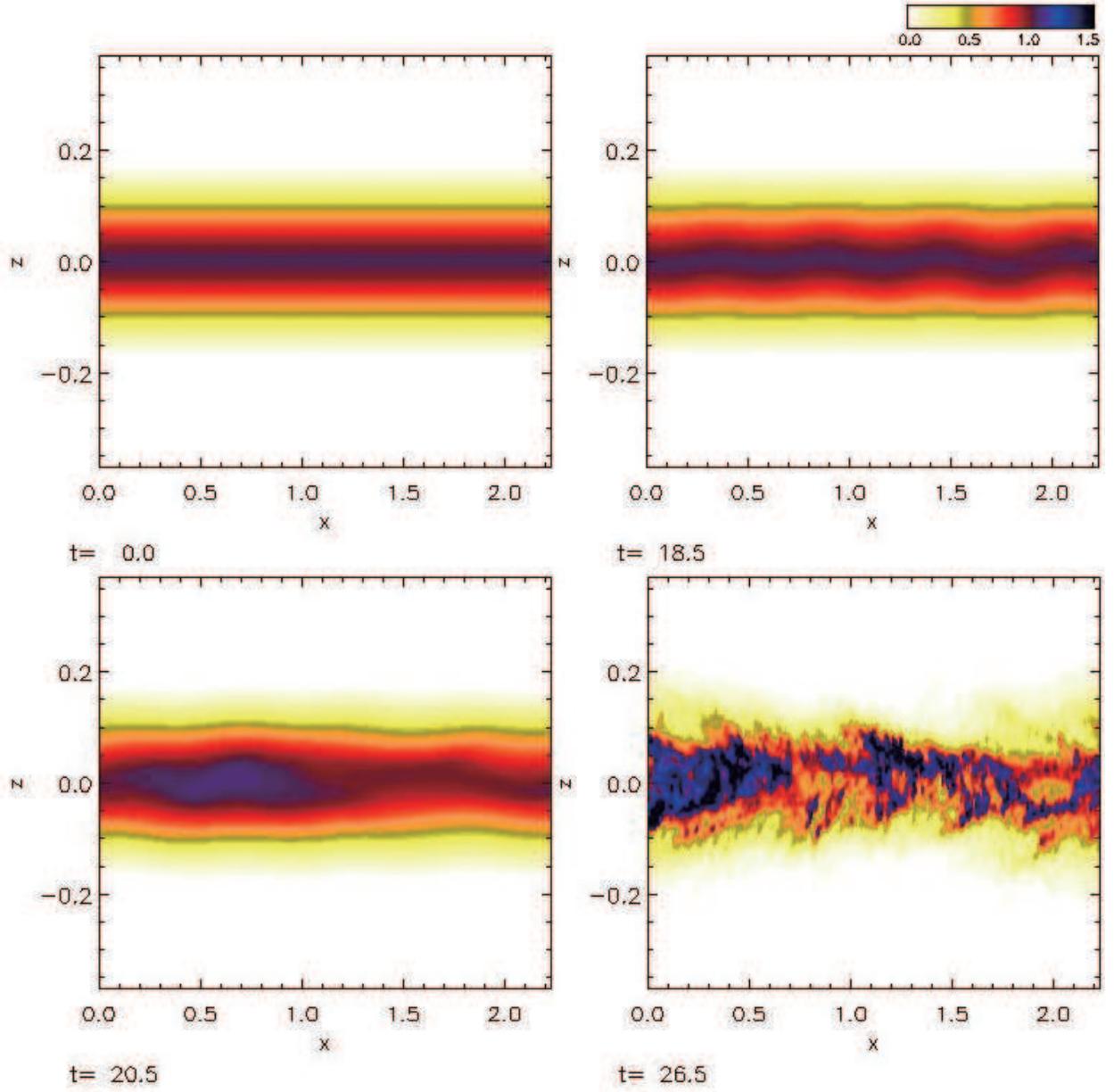}
\caption{
Snapshots of dust density at times $t \Omega_K=0, 18.5, 20.5$ and 26.5,
in the case where $\tau_f \Omega_K=1$ and  $\rho_d(0)/\rho_g =1$.
\label{densitygt0}}
\end{figure}

The r.m.s. of a vertical dust velocity for $\tau_f \Omega_K=1$
is two orders magnitude smaller than that for $\tau_f \Omega_K=10^{-3}$
as seen in Figure \ref{t-wdrms}.
The reason would be that the coupling between gas and dust is weak.
As shown by \citet{johansen07a}, dust concentrates 
by the streaming instability  after the flow becomes turbulent.
That is also seen in our simulation at $t \Omega_K=26.5$ as shown
in Figure \ref{densitygt0}.

In our calculation, the dust does not settle down 
because the gravity of central star is neglected.
In reality, the dust would settle down before the start of 
the shear instability
for the parameter, $\rho_d(0)/\rho_g=1, h_d/ z_d =0.5$ and
$\tau_f \Omega_K=1$ because the settling time 
$1/(\tau_f \Omega_K^2) \sim 1/\Omega_K$ is shorter than
the growth time of the turbulence. 
As dust settling proceeds, the shear instability occurs
when the growth time of shear instability becomes shorter than
the dust settling time \citep{johansen06b, johansen07a}.

\section{ENERGY SOURCES OF INSTABILITY}
In this section, we analyze the energy source of the instabilities 
presented in \S 3.
The energy budget for the instability can be estimated from 
distributions of density and 
velocities obtained by our simulations shown in \S3.

\subsection{Linearization}
We linearize equations (\ref{eqn:ba1})--(\ref{eqn:ba4}) as follows:
\begin{equation}
	  \DP{u'_g}{x} +  \DP{w'_g}{z} =0,
	 \label{eqn:pge1}
 \end{equation}
\begin{equation}
	 \DP{u'_g}{t} + \bar{u}_g \DP{u'_g}{x} + w'_g \DP{ \bar{u}_g}{z}
      = - \frac{1}{\rho_g} \DP{P'}{x}  
        +2 \Omega_K  v'_g - \frac{1}{\tau_f \rho_g} \bar{\rho}_d(u'_g - u'_d)
	- \frac{1}{\tau_f \rho_g} \rho'_d(\bar{u}_g - \bar{u}_d),
	\label{eqn:pge2}
\end{equation} 
 \begin{equation}
	  \DP{v'_g}{t} + \bar{u}_g \DP{v'_g}{x} + w'_g \DP{\bar{v}_g}{z}
       =  - \frac{1}{2} \Omega_K u'_g - \frac{1}{\tau_f \rho_g} 
          \bar{\rho}_d (v'_g - v'_d)
	- \frac{1}{\tau_f \rho_g} \rho'_d (\bar{v}_g - \bar{v}_d),
	 \label{eqn:pge3}
 \end{equation}
 \begin{equation}
	  \DP{w'_g}{t} + \bar{u}_g \DP{w'_g}{x} 
       = - \DP{P'}{z} 
	- \frac{1}{\tau_f \rho_g}  \bar{\rho}_d(w'_g - w'_d).
         \label{eqn:pge4}
 \end{equation}
 \begin{equation}
	 \DP{\rho'_d}{t}  +   \bar{\rho}_d  \DP{u'_d}{x} 
	+ \bar{u}_d \DP{ \rho'_d }{x} 
	+ \bar{\rho}_d \DP{ w'_d}{z}
        + \DP{\bar{\rho}_d}{z} w'_d  =0,
	 \label{eqn:pde1}
 \end{equation}
\begin{equation}
	 \DP{u'_d}{t} + \bar{u}_d \DP{u'_d}{x} + w'_d \DP{\bar{u}_d}{z}
      = 2 \Omega_K v'_d - \frac{1}{\tau_f } (u'_d - u'_g),
	\label{eqn:pde2}
\end{equation} 
 \begin{equation}
	  \DP{v'_d}{t} + \bar{u}_d \DP{v'_d}{x} 
	+ w'_d \DP{\bar{v}_d}{z}
       = - \frac{1}{2} \Omega_K u'_d - \frac{1}{\tau_f } (v'_d - v'_g),
	 \label{eqn:pde3}
 \end{equation}
 \begin{equation}
	  \DP{w'_d}{t} + \bar{u}_d \DP{w'_d}{x}
       = - \frac{1}{\tau_f} (w'_d - w'_g).
	 \label{eqn:pde4}
 \end{equation}
Assuming $f' \propto e^{i(kx-\omega t)}$ ($k$ denotes the radial wave 
number, and $\omega= \omega_R + i \omega_I$ denotes the complex frequency),
equations (\ref{eqn:pge1})--(\ref{eqn:pde4}) are written as
\begin{equation}
	  i k u'_g +  \DD{w'_g}{z} =0,
	 \label{eqn:kpge1}
 \end{equation}
\begin{equation}
	 - i \tilde{\omega}_g u'_g + w'_g \DD{ \bar{u}_g}{z}
      = - i k \frac{1}{\rho_g}P' +2 \Omega_K v'_g 
             - \frac{1}{\tau_f \rho_g} \bar{\rho}_d(u'_g - u'_d)
	- \frac{1}{\tau_f \rho_g} \rho'_d(\bar{u}_g - \bar{u}_d),
	\label{eqn:kpge2}
\end{equation} 
 \begin{equation}
	  - i \tilde{\omega}_g v'_g+ w'_g \DD{\bar{v}_g}{z}
       =  - \frac{1}{2} \Omega_K u'_g 
          - \frac{1}{\tau_f \rho_g} \bar{\rho}_d (v'_g - v'_d)
	- \frac{1}{\tau_f \rho_g} \rho'_d (\bar{v}_g - \bar{v}_d),
	 \label{eqn:kpge3}
 \end{equation}
 \begin{equation}
	  - i \tilde{\omega}_g w'_g 
       = - \DD{P'}{z} 
	- \frac{1}{\tau_f \rho_g}  \bar{\rho}_d(w'_g - w'_d).
         \label{eqn:kpge4}
 \end{equation}
 \begin{equation}
	 - i \tilde{\omega}_d \rho'_d  +   i k \bar{\rho}_d u'_d 
	+ \bar{\rho}_d \DD{ w'_d}{z} + \DD{\bar{\rho}_d}{z} w'_d =0,
	 \label{eqn:kpde1}
 \end{equation}
\begin{equation}
	 - i \tilde{\omega}_d u'_d + w'_d \DD{\bar{u}_d}{z}
      = 2 \Omega_K v'_d - \frac{1}{\tau_f } (u'_d - u'_g),
	\label{eqn:kpde2}
\end{equation} 
 \begin{equation}
	   - i \tilde{\omega}_d v'_d 
	+ w'_d \DD{\bar{v}_d}{z}
       = - \frac{1}{2} \Omega_K u'_d - \frac{1}{\tau_f } (v'_d - v'_g),
	 \label{eqn:kpde3}
 \end{equation}
 \begin{equation}
	   - i \tilde{\omega}_d w'_d
       = - \frac{1}{\tau_f } (w'_d - w'_g),
	 \label{eqn:kpde4}
 \end{equation}
where $ \tilde{\omega}_g = \omega - k \bar{u}_g$ and
$ \tilde{\omega}_d = \omega - k \bar{u}_d$.

\subsection{Energy Equations}

We consider the radial energy budget for gas.
Multiplying equation (\ref{eqn:kpge2}) by $u'^*_g$ yields
\begin{eqnarray}
	 - i \tilde{\omega}_g |u'_g|^2 
       &= & - \DD{ \bar{u}_g}{z} w'_g u'^*_g 
           - i k \frac{1}{\rho_g} P' u'^*_g +2 \Omega_K v'_g u'^*_g 
	- \frac{1}{\tau_f \rho_g } \bar{\rho}_d(u'_g - u'_d) u'^*_g  \nonumber \\ 
       &  & + \frac{1}{\tau_f \rho_g} (\bar{u}_g - \bar{u}_d) \rho'_d u'^*_g,
\end{eqnarray} 
where the superscript ${}^*$ denotes the complex conjugate.
Taking the real part yields
\begin{eqnarray}
	  \omega_I |u'_g|^2 
       &= & - \DD{ \bar{u}_g}{z} \Re[w'_g u'^*_g] 
        + k \frac{1}{\rho_g} \Im[P' u'^*_g]  + 2 \Omega_K\Re[v'_g u'^*_g] 
	- \frac{1}{\tau_f \rho_g }\bar{\rho}_d 
           \Re [(u'_g - u'_d) u'^*_g] \nonumber \\ 
       &  &  - \frac{1}{\tau_f \rho_g} (\bar{u}_g - \bar{u}_d)  
           \Re \left[\rho'_d u'^*_g \right].
	\label{eqn:reneg}
\end{eqnarray} 
We perform the similar manipulations for azimuthal and vertical directions,
\begin{eqnarray}
	  \omega_I |v'_g|^2 
       &= & - \DD{ \bar{v}_g}{z} \Re[w'_g v'^*_g] 
        + 2 \Omega_K \Re[u'_g v'^*_g] 
	- \frac{1}{\tau_f \rho_g}\bar{\rho}_d 
        \Re [(v'_g - v'_d) v'^*_g] \nonumber \\ 
       &  &  - \frac{1}{\tau_f \rho_g} (\bar{v}_g - \bar{v}_d)  
           \Re \left[\rho'_d v'^*_g \right],
	\label{eqn:teneg}
\end{eqnarray} 
\begin{eqnarray}
	  \omega_I |w'_g|^2 
       = -\frac{1}{\rho_g} \Re\left[\DD{P}{z}' w'^*_g\right]
           - \frac{1}{\tau_f \rho_g}\bar{\rho}_d 
        \Re [(w'_g - w'_d) w'^*_g] 
              - \frac{1}{\tau_f \rho_g} (\bar{w}_g - \bar{w}_d)  
           \Re \left[\rho'_d w'^*_g \right].
	\label{eqn:zeneg}
\end{eqnarray} 
Similarly, we consider the energy budgets for dust.
\begin{eqnarray}
	  \omega_I |u'_d|^2 
       &= & - \DD{ \bar{u}_d}{z} \Re[w'_d u'^*_d] 
       + 2 \Omega_K \Re[v'_d u'^*_d] 
	- \frac{1}{\tau_f} \Re [(u'_d - u'_g) u'^*_d],
	\label{eqn:rened}
\end{eqnarray} 
\begin{eqnarray}
	  \omega_I |v'_d|^2 
       &= & - \DD{ \bar{v}_d}{z} \Re[w'_d v'^*_d] 
       - \frac{1}{2} \Omega_K \Re[u'_d v'^*_d] 
	- \frac{1}{\tau_f} \Re [(v'_d - v'_g) v'^*_d],
	\label{eqn:tened}
\end{eqnarray} 
\begin{eqnarray}
	  \omega_I |w'_d|^2 
       &= & 	- \frac{1}{\tau_f} \Re [(w'_d - w'_g) w'^*_d].
	\label{eqn:zened}
\end{eqnarray} 

\begin{figure}
\plotone{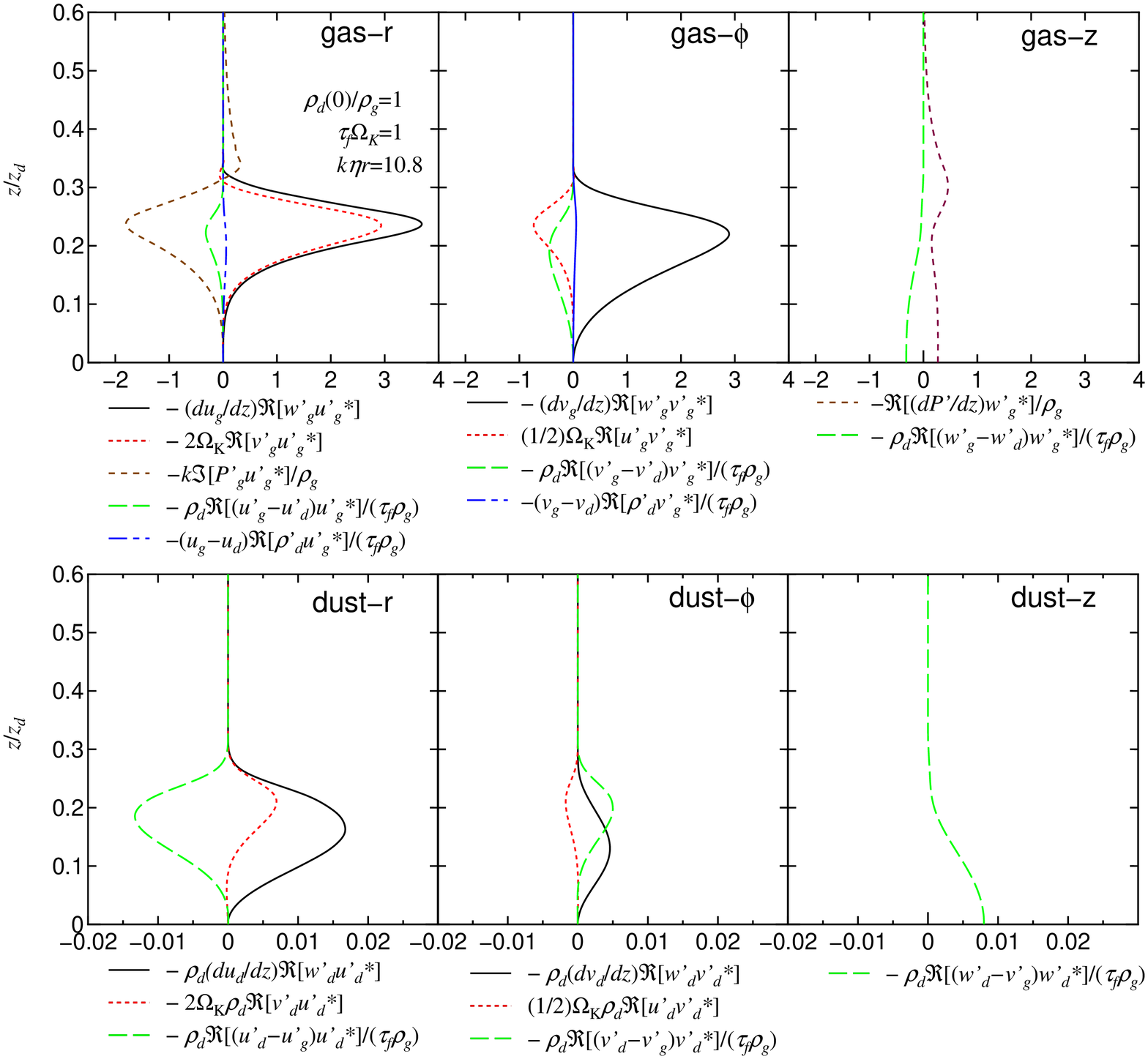}
\caption{
Each term in the right side of energy equations 
(\ref{eqn:reneg})--(\ref{eqn:zened}),
in the case where $k \eta r=10.8 $, $\tau_f \Omega_K=1$ and 
$\rho_d(0)/\rho_g =1$.
\label{ener1t0}}
\end{figure}

\begin{figure}
\plotone{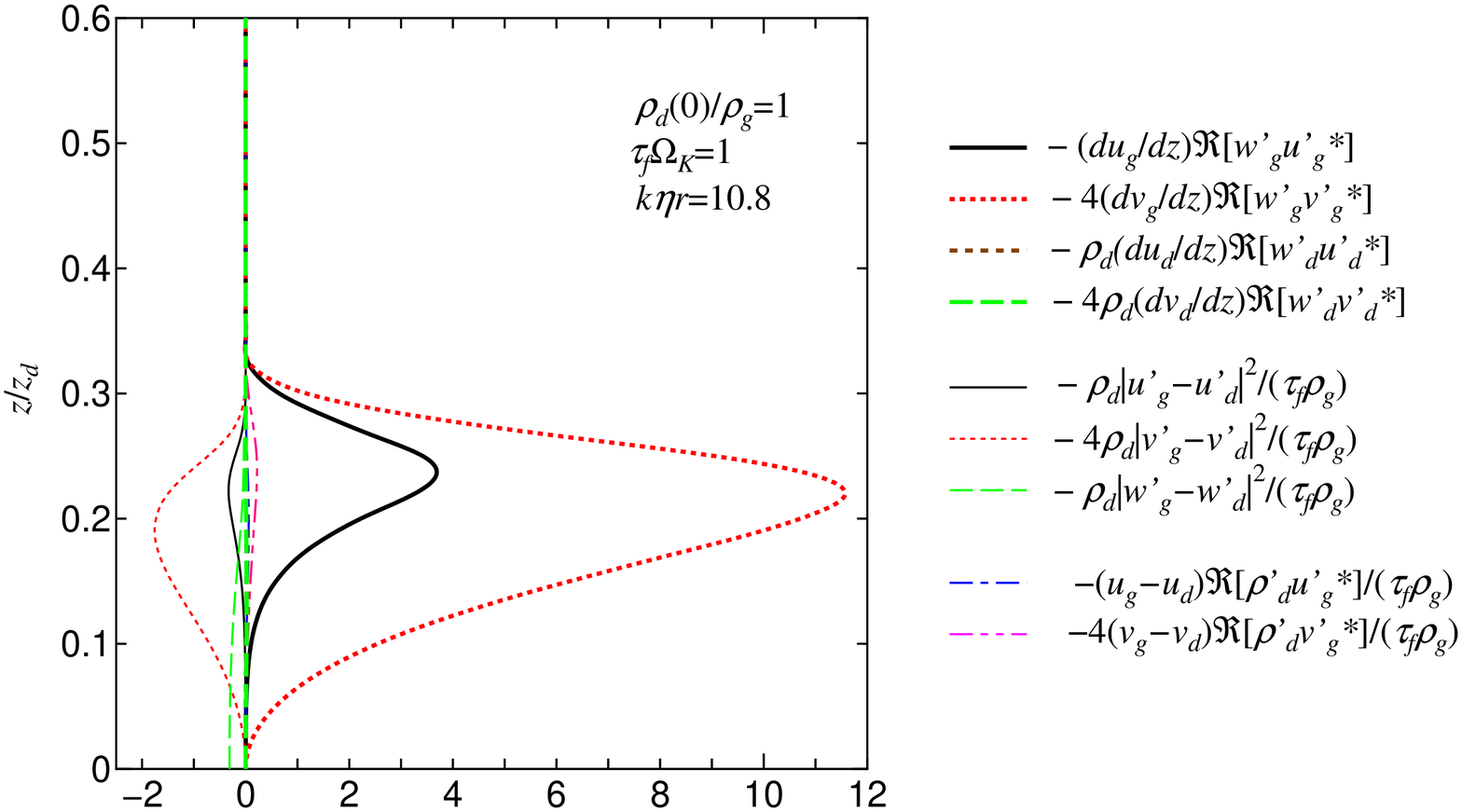}
\caption{
Each term in the right side of energy equation
(\ref{eqn:total_ene}),
in the case where $k \eta r=10.8 $, $\tau_f \Omega_K=1$ and 
$\rho_d(0)/\rho_g =1$.
\label{total_ener1t0}}
\end{figure}

Here, let us see the total energy budget.
Multiplying equations (\ref{eqn:reneg})--(\ref{eqn:zeneg})  by  $\rho_g$
and equations (\ref{eqn:rened})--(\ref{eqn:zened}) by $\bar{\rho}_d$,
we sum them.
Note that equations (\ref{eqn:teneg}), (\ref{eqn:tened}) is 
multiplied by 4 in order to eliminate Epicyclic terms.
\begin{eqnarray}
       &  &   \omega_I \{ \rho_g ( |u'_g|^2 +4 |u'_g|^2 + |u'_g|^2 )
           + \bar{\rho}_d ( |u'_d|^2 +4 |u'_d|^2 + |u'_d|^2 ) \}  \nonumber \\ 
      & = & - \left\{ \DD{ \bar{u}_g}{z} \Re[w'_g u'^*_g] 
             +  \DD{ \bar{u}_d}{z} \Re[w'_d u'^*_d] 
             +  4 \DD{ \bar{v}_g}{z} \Re[w'_g v'^*_g] 
             +  4 \DD{ \bar{v}_d}{z} \Re[w'_d v'^*_d] \right\}  \nonumber \\ 
      & &  - \Re\left[ i k P' u'^*_g + \DD{P'}{z} w'^*_g \right]
	- \frac{\bar{\rho}_d }{\tau_f } \left\{  |u'_g - u'_d|^2 
            + 4  |v'_g - v'_d|^2 +  |w'_g - w'_d|^2 \right\} \nonumber \\ 
       &  &  - \frac{1}{\tau_f }  \left\{ 
             (\bar{u}_g - \bar{u}_d)  \Re \left[\rho'_d u'^*_g \right]
          +4 (\bar{v}_g - \bar{v}_d)  \Re \left[\rho'_d v'^*_g \right]  \right\}.
	\label{eqn:total_ene}
\end{eqnarray} 
The equation (\ref{eqn:total_ene}) is integrated in the $z$-direction
in order to examine the total energy budget.
We focus on the pressure term of the second term of 
the right hand in the equation (\ref{eqn:total_ene}).
\begin{equation}
  \int \DD{P'}{z} w'^*_g dz = - \int P' \left( \DD{w'_g}{z}\right)^* dz
         = - \int P' iku'^*_gdz,
\end{equation}
where the partial integration and the periodic boundary condition
are used in the first equality, and the equation (\ref{eqn:kpge1})
is used in the second equality.
We have 
\begin{equation}
  \int \left( P' iku'^*_g + \DD{P'}{z} w'^*_g \right)dz  = 0.
\end{equation}
Thus, the total energy budget of the system is determined 
by the shear terms, the friction terms due to the fluctuation 
of the dust density, and the frictional dissipation terms;
the total energy does not change by the pressure terms and epicyclic terms,
and they merely exchange energy among $x$, $y$, and $z$ directions.
Figure \ref{ener1t0} shows the each term in energy equations 
(\ref{eqn:reneg})--(\ref{eqn:zened})
and Figure \ref{total_ener1t0} shows the each term in total energy equation 
(\ref{eqn:total_ene})
for $k \eta r= 10.8$ at which the growth rate has 
maximum value for  $\tau_f \Omega_K=1$ and $\rho_d(0)/\rho_g =1$.
Then the shear of radial and azimuthal velocities of gas 
is a main energy source.

\begin{figure}
\plotone{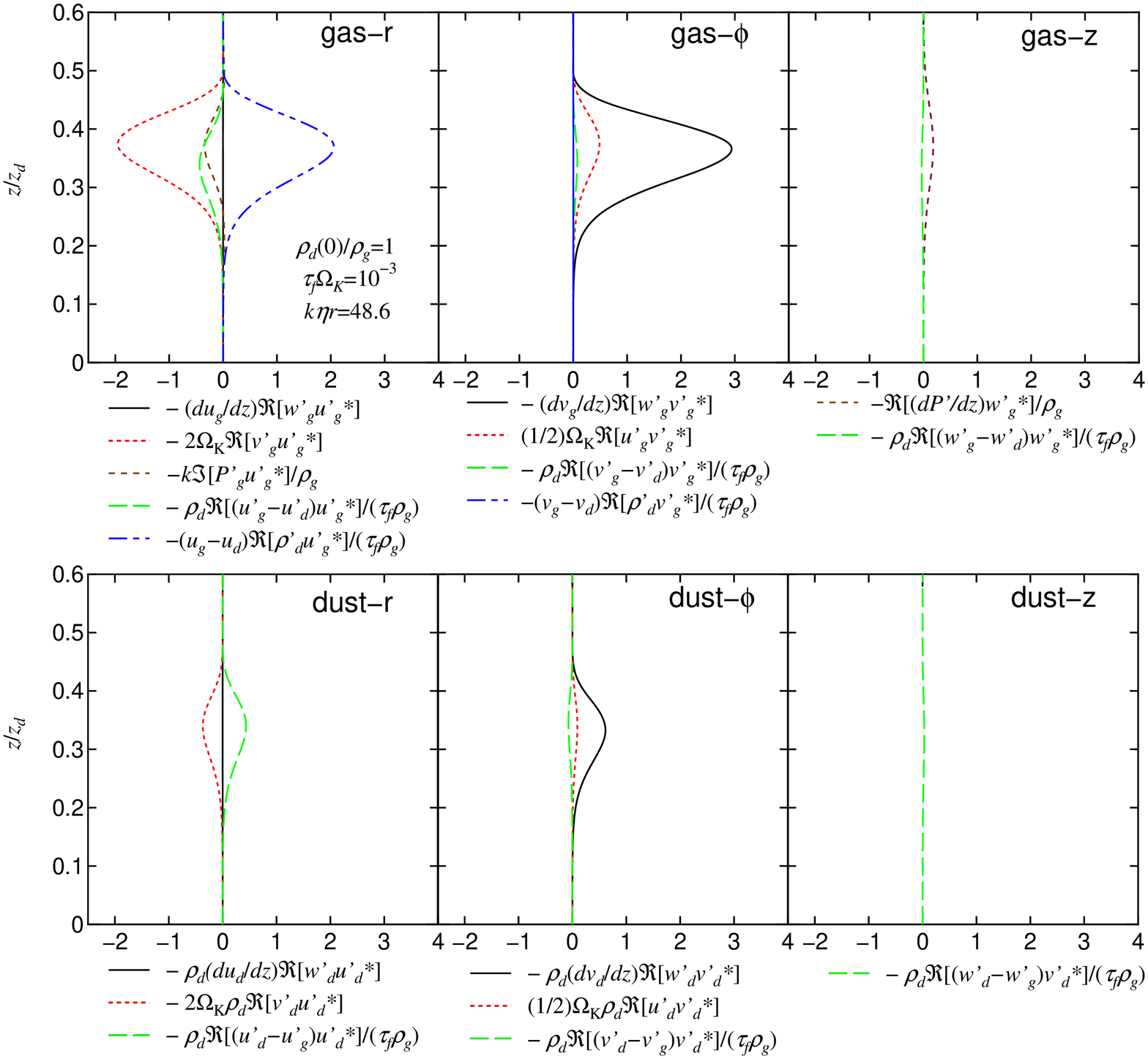}
\caption{
Each term in the right side of energy equations
(\ref{eqn:reneg})--(\ref{eqn:zened}),
in the case where $k \eta r= 48.6$, $\tau_f \Omega_K=10^{-3}$ and
 $\rho_d(0)/\rho_g =1$.
\label{ener1t3}}
\end{figure}

\begin{figure}
\plotone{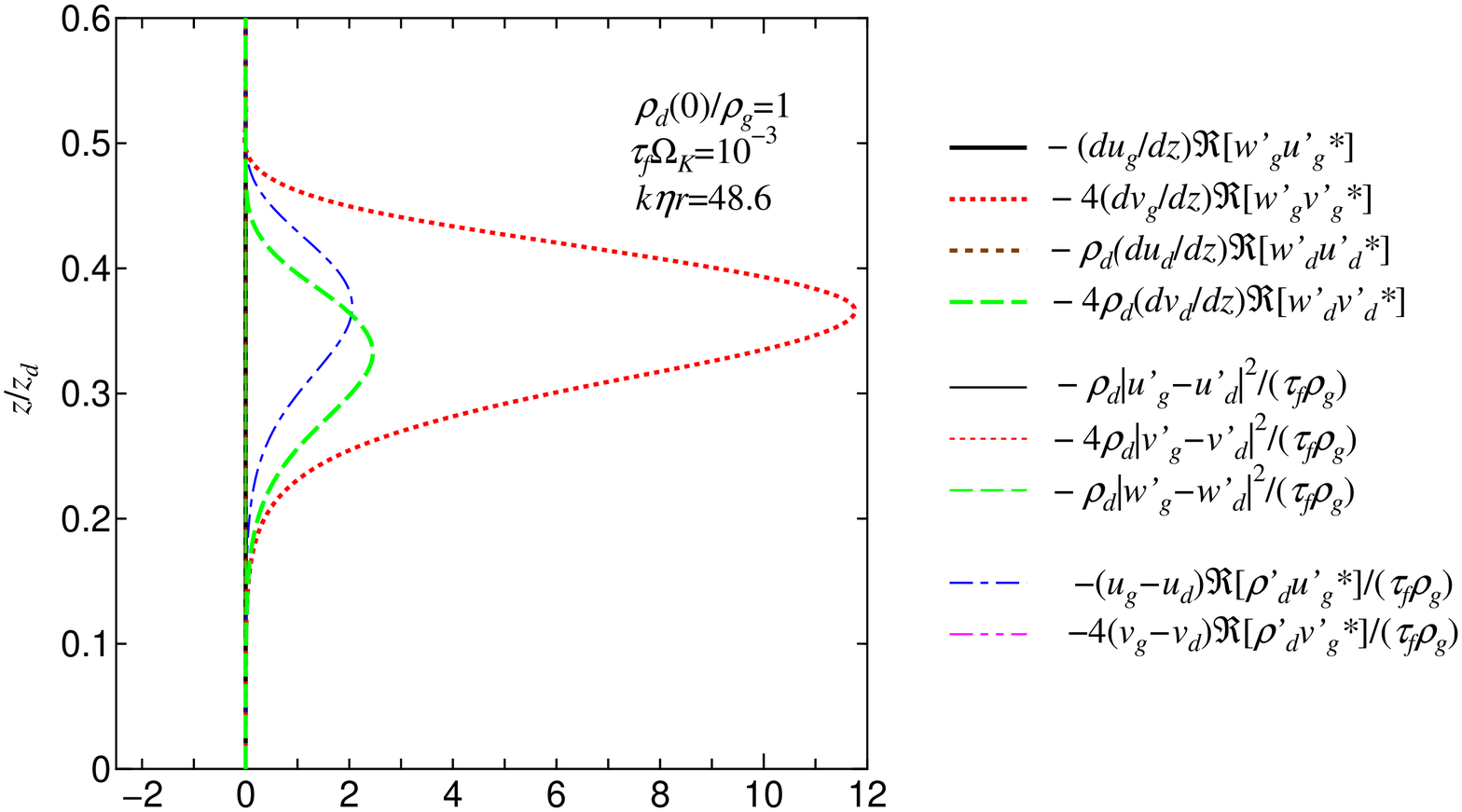}
\caption{
Each term in the right side of energy equation
(\ref{eqn:total_ene}),
in the case where $k \eta r= 48.6$, $\tau_f \Omega_K=10^{-3}$ and
 $\rho_d(0)/\rho_g =1$.
\label{total_ener1t3}}
\end{figure}

Figure \ref{ener1t3} shows the each term in energy equations 
(\ref{eqn:reneg})--(\ref{eqn:zened})
and Figure \ref{total_ener1t3} shows the each term in total energy equation 
(\ref{eqn:total_ene})
for $k \eta r= 48.6$ at which the growth rate has 
maximum value for   $\tau_f \Omega_K=10^{-3}$ and $\rho_d(0)/\rho_g =1$.
The shears of azimuthal velocities $d\bar{v}_g/dz$ and $d\bar{v}_d/dz$ 
are main energy sources.
In addition, the radial relative velocity between dust and gas, coupled
with the density fluctuations, are also an important energy source.

\begin{figure}
\plotone{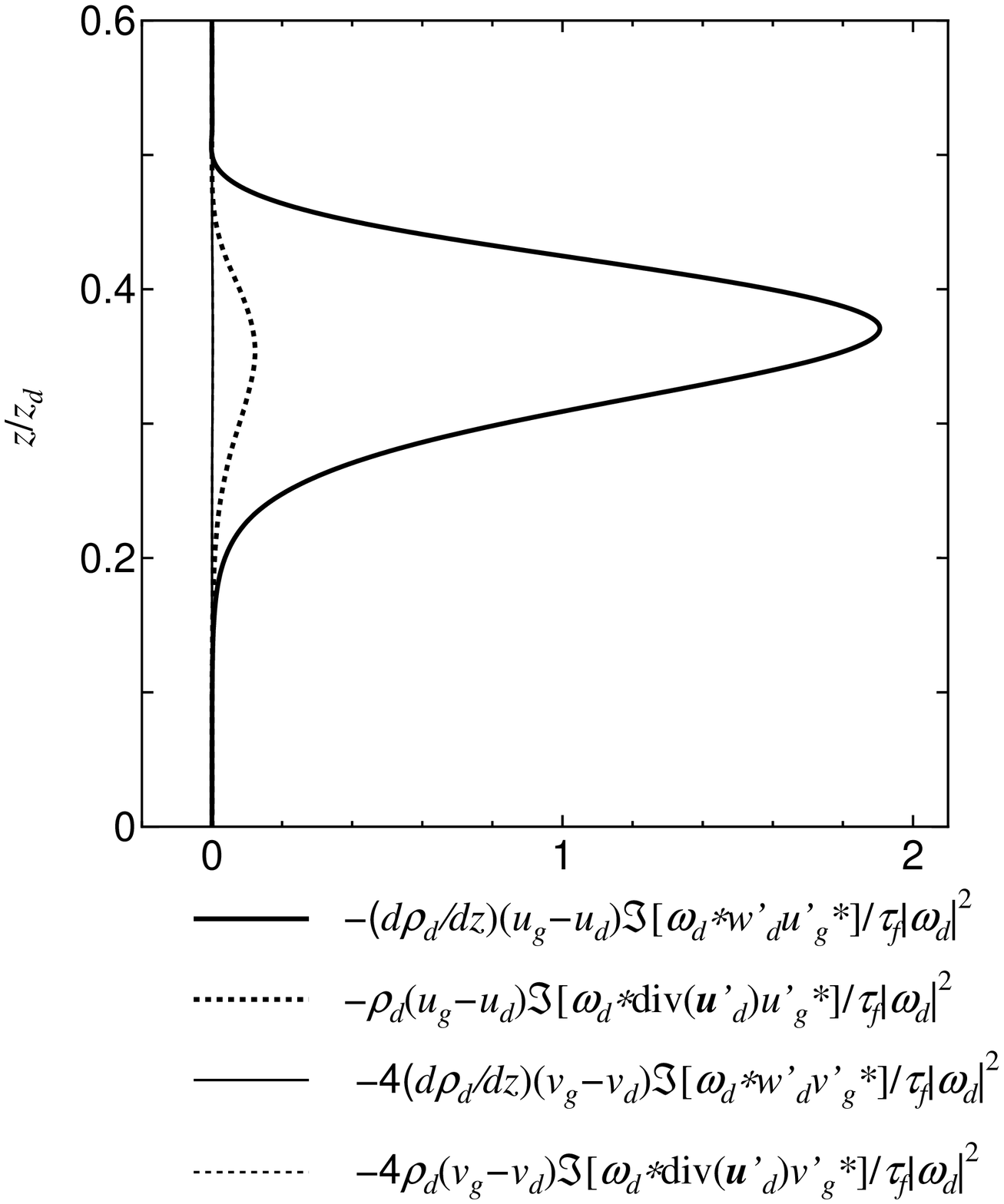}
\caption{
Each term in the right side of energy equation (\ref{eqn:pedust}),
in the case where $k \eta r= 48.6$, $\tau_f \Omega_K=10^{-3}$ and
 $\rho_d(0)/\rho_g =1$.
\label{ener1t3dz}}
\end{figure}

In energy equation (\ref{eqn:total_ene}), the last term 
$ - \frac{1}{\tau_f }  \left\{ 
(\bar{u}_g - \bar{u}_d)  \Re \left[\rho'_d u'^*_g \right]
+4 (\bar{v}_g - \bar{v}_d)  \Re \left[\rho'_d v'^*_g \right]  \right\}$
shows an instability powered by  relative velocity of gas and dust,
coupled with Eulerian dust density fluctuation.
Substituting the linearized continuity equation for dust (\ref{eqn:kpde1}),
\begin{equation}
	 \rho'_d  = - i \frac{\tilde{\omega}^*_d}{|\tilde{\omega}_d|^2} 
	 \left(   i k \bar{\rho}_d u'_d 
	+ \bar{\rho}_d \DD{ w'_d}{z} + \DD{\bar{\rho}_d}{z}  w'_d  \right),
        \label{eqn:coedust}
 \end{equation}
this term is divided into parts
\begin{eqnarray}
&&  - \frac{1}{\tau_f} \left\{ (\bar{u}_g - \bar{u}_d)  
           \Re \left[\rho'_d u'^*_g \right] 
          +4  (\bar{v}_g - \bar{v}_d)  
           \Re \left[\rho'_d v'^*_g \right]  \right\}         \nonumber \\
       &= & - \frac{1}{\tau_f} \frac{1}{|\tilde{\omega}_d|^2}  
         \bar{\rho}_d  \left\{ 
        (\bar{u}_g - \bar{u}_d)  \Im \left[\tilde{\omega}^*_d
	\left( i k u'_d + \DD{w'_d}{z} \right) u'^*_g \right]
       +4 (\bar{v}_g - \bar{v}_d)  \Im \left[\tilde{\omega}^*_d
	\left( i k u'_d + \DD{w'_d}{z} \right) v'^*_g \right] \right\} 
        \nonumber \\
       &  & - \frac{1}{\tau_f} \frac{1}{|\tilde{\omega}_d|^2}  
        \DD{\bar{\rho}_d }{z} \left\{
        (\bar{u}_g - \bar{u}_d) \Im[ \tilde{\omega}^*_d w'_d  u'^*_g ]
        +4  (\bar{v}_g - \bar{v}_d) \Im[ \tilde{\omega}^*_d w'_d  v'^*_g ]
         \right\}.
        \label{eqn:pedust}
\end{eqnarray} 
The first term in the right hand of equation (\ref{eqn:pedust})
presumably corresponds to the streaming instability
addressed by \citet{youdin05}, \citet{youdin07} and  \citet{johansen07a}.
The divergence of the dust velocity $\left( i k u'_d + \DD{w'_d}{z} \right)$ 
denotes that the increase of the fluctuation of dust density
is related to the streaming instability.
If $\rho_g \gg \rho_d$ and $\tau_f \Omega_K \ll 1$, 
the divergence of the dust velocity
approaches zero owing to the incompressibility of gas
and a small relative velocity between dust and gas.
Thus, if dust size is small, the contribution of the first 
term of the right hand side of equation (\ref{eqn:pedust}) to the instability 
becomes small.
This is consistent with the results of \citet{youdin05} and \citet{youdin07}.
The second term on the right hand of equation (\ref{eqn:pedust})
also denotes the power due to the fluctuation of dust density.
However, the dust density fluctuation is produced by the vertical advection
under the condition with an initial vertical dust density gradient.
The fluctuation of dust density is the cause of instability 
in either term of the equation (\ref{eqn:pedust}).
The first term arises due to the Lagrangian density fluctuation.
On the other hand, the cause of the density fluctuation is 
advection in the second term.
The growth rate of instability $\omega_I$ and the phase velocity
$v_p = \omega_R / k$ are estimated from the numerical result.
Figure \ref{ener1t3dz} shows each term on the left hand of
 equation (\ref{eqn:pedust}) for 
$\tau_f \Omega_K=10^{-3}$, $\rho_d(0)/\rho_g=1$ and $h_d/ z_d =0.5$.
Obviously, the energy is gained from the second term 
associated with Eulerian density fluctuation.

Because \citet{youdin05} and  \citet{youdin07} have assumed
that dust density is constant, 
the instability caused by the second term in the equation (\ref{eqn:pedust})
has not been seen.
In a density stratified layer, if the dust size is small, 
the instability which obtains the energy through the second term 
in addition to the shear terms occurs.

Let us see the relation between the two-fluid instability and 
the baroclinic instability stated in \citet{ishitsu03}.
One fluid approximation assumes infinitesimal friction time 
$\tau_f \rightarrow 0$.
In order for the friction term in equations (\ref{eqn:ba2}) 
and (\ref{eqn:ba4}) to be finite,
we have $ \ibm{u}_g \rightarrow \ibm{u}_d$.
Eliminating the friction terms by adding equation (\ref{eqn:ba2})
$\times (\rho_g / \rho)$ and equation (\ref{eqn:ba4}) $\times (\rho_d / \rho)$ 
where $\rho \equiv  \rho_g +\rho_d$, we have
\begin{equation}
	 \DP{\ibm{u}}{t} + ( \ibm{u} \cdot \nabla_2 ) \ibm{u}
      = - \frac{1}{ \rho } \nabla_2 P + 2 \ibm{u} \times \ibm{\Omega}_K
       + \frac{3}{2} u \Omega_K \hat{\ibm{y}},
	\label{eqn:1meq}
\end{equation} 
where $\ibm{u} \equiv \ibm{u}_g = \ibm{u}_d$.
Linearizing the pressure term in the equation (\ref{eqn:1meq}) yields
\begin{equation}
-\frac{1}{\bar{\rho} + \rho'} \DP{(\bar{P} + P')}{r} 
\approx - \frac{1}{\bar{\rho}}\DP{P'}{r}
+ \frac{1}{\bar{\rho}^2} \DP{\bar{P}}{r} \rho'.
	\label{eqn:1pg}
\end{equation}
The second term in the right hand is the buoyancy term, which is 
the cause of the baroclinic instability for the one-fluid approximation
 \citep{ishitsu02}.
Note that we assume the gas to be incompressible.
In the one-fluid model, the fluid of dust and gas mixture
is also incompressible because 
$\nabla \cdot \ibm{u} = \nabla \cdot \ibm{u}_g=0$.
Thus, the dust density perturbation arises not from the 
compression but from the advection; the latter can have non-zero value
if $\nabla \bar{\rho} \neq 0$.
Adding equation (\ref{eqn:pge2}) by $\rho_g/\bar{\rho}$ 
and equation (\ref{eqn:pde2}) by 
$\bar{\rho}_d/\bar{\rho}$, we have
\begin{eqnarray}
& & \DP{}{t} \left( \frac{ \rho_g u'_g 
+ \bar{\rho}_d u'_d} { \bar{\rho}} \right)
 + \frac{\rho_g \bar{u}_g}{\bar{\rho}} \DP{u'_g}{x}
 + \frac{\bar{\rho_d} \bar{u}_d}{\bar{\rho}} \DP{u'_d}{x}
+ \frac{ \rho_g w'_g}{\bar{\rho}}  \DP{\bar{u}_g }{z} 
+ \frac{ \bar{\rho_d} w'_g}{\bar{\rho}}  \DP{\bar{u}_d }{z} \nonumber \\ 
 &=& 
-\frac{1}{\bar{\rho}} \DP{P'}{z} + 2 \Omega_K 
\frac{\rho_g v'_g + \bar{\rho}_d v'_d}{\bar{\rho}}
-\frac{\rho'_d}{\tau_f \bar{\rho}} ( \bar{u}_g-\bar{u}_d).
\label{eqn:pgde}
\end{eqnarray}
It is easily seen that the left hand side and the Coriolis force term 
in the right hand side become those of the linearized equation 
of equations (\ref{eqn:1meq})
in the limit  $\ibm{u}_g \rightarrow \ibm{u}_d$. 
We see the frictional term in the right hand of the equation (\ref{eqn:pgde})
is written
\begin{eqnarray}
   - \frac{1}{\tau_f \bar{\rho}} (\bar{u}_g -\bar{u}_d) \rho'_d
           & = &  - \frac{1}{\tau_f \bar{\rho}} 
             \frac{ 2 ( \rho_g+\bar{\rho}_d   ) \rho_g \tau_f \Omega_K}
                  { (\rho_g + \bar{\rho}_d )^2 + ( \rho_g \tau_f \Omega_K)^2}
                 \eta v_K \rho'_d \nonumber \\ 
            & = &  \frac{1}{\bar{\rho}} \frac{  \rho_g + \bar{\rho}_d }
                  { (\rho_g + \bar{\rho}_d )^2 
                 + ( \rho_g \tau_f \Omega_K)^2} \DP{P}{r} \rho'_d, 
\end{eqnarray}
where we use equations (\ref{eqn:eta}), (\ref{eqn:nueq1}),
 and (\ref{eqn:nueq3}).
Limiting to the one-fluid leads
\begin{eqnarray}
     - \frac{1}{\tau_f \rho_g} (\bar{u}_g -\bar{u}_d) \rho'_d
            \rightarrow   \frac{1}
                  { \bar{\rho}^2 }
                     \DP{P}{r} \rho'_d  \;\;\; \mbox{as} \;\;\; 
                   \tau_f \rightarrow 0.
\end{eqnarray}
This term corresponds to the second term in the right hand 
of the equation (\ref{eqn:1pg}).
However, two-fluid instability is different from the 
baroclinic instability with regards that
the baroclinic instability has no axis-symmetric mode.

\section{Conclusions}
We performed the two-fluid of gas and dust, two-dimensional simulations
in the dust layer of a protoplanetary disk.
For $\tau_f \Omega_K=10^{-3}$,
the numerical simulations show the rapid growth instability
induced mainly by the vertical shear of azimuthal velocity, and 
additionally the relative motion between dust and gas
coupled with the dust density fluctuation due to advection
if the dust density distribution has significant gradient,
$ |\frac{d\rho_d}{dz}| \gtrsim \rho_d(0)/\eta r$.
The streaming instability stated by \citet{youdin07},
which is caused by the relative motion of dust and gas coupled
 with the Lagrangian dust density fluctuation,
has the small growth rate of the instability 
if the dust size is smaller than several centimeters. 
On the other hand, the instability 
powered by the vertical shear of the azimuthal velocity, and 
additionally by the relative velocity of dust and gas coupled with
dust density fluctuation due to advection
shown in this work has the growth rate $~ \Omega_K$
even if the dust size is small. 

The density fluctuations grows due to the streaming instability
if the initial dust density is constant, which is accompanied by
the concentrations of dust density.
However, if the initial dust density is not constant,
the instability related by the vertical dust density gradient occurs.
The latter instability diffuses the dust rather than concentrates that.
This suggests that the maximum density does not always  
increase from the initial value.

Additionally, for $\tau_f \Omega_K=1$, the simulations 
shows the vertical shear of the radial flow plays the important role.
After the flow becomes a turbulent state due to the shear instability, 
the dust concentrations are induced by the streaming instability
because large relative motion between dust and gas is permissible
due to loose coupling of dust and gas.

\citet{chiang08} and \citet{barranco09} have performed
the one-fluid, 3D simulations of the shear instability 
and presented that the conditions of the transition to turbulence
depends not only on the Richardson number but also 
on the initial perturbations.
The dependency of the shear instability on initial conditions 
is explained as follows.
The growth rate of the instability depends on the radial and
the azimuthal wave numbers $k_r$ and $k_{\phi}$ if there is 
no the radial shear such as the Kepler shear.
The unstable region in the Fourier space is restricted to
a small value of  $k_{r}$ (see  figure 1 in \citet{ishitsu03}).
In addition, the shear instability does not have 
the axis-symmetric unstable mode,
that is, $k_{\phi}=0$ mode. 
If there is radial shear due to the Kepler motion, 
the radial wave number increases due to shear-stretching as time passes 
(see equation (43) in \citet{ishitsu03}),
\begin{equation}
   k_r(t) = k_r(0) + \frac{3}{2} k_{\phi} t \Omega_K.
\end{equation}
The perturbation can grow only when its wave number passes 
the unstable region in the Fourier space.
The flow can transit into turbulence
due to the transient amplification if the initial perturbation 
is large. On the other hand, the flow cannot transit into turbulence 
for small values of initial perturbations.

However, the instability of two-fluid shown in this work
has the axis-symmetric unstable mode.
As a result, the stabilization caused by the increase of the 
azimuthal wave number due to the radial shear is not effective.
We expect that the instability occurs in the radial direction, 
and then the perturbation with small azimuthal wave number grows.

\citet{chiang08} estimates the critical dust surface density
as the condition that the shear instability does not occur 
if there is not a global turbulence such as  MRI.
This estimation is derived from one-fluid simulations.
However, in two-fluid of gas and dust,
the instability due to the vertical dust density gradient 
and the relative motion between gas and dust occurs.
The flow can transit into turbulence
even if the disk has the critical surface density estimated by \citet{chiang08}.
Thus, the instability induced by the dust density gradient 
may preclude the planetesimal formation due to the gravitational instability.

In the field of the meteoritics, 1 Myr time-lag between
the formations of  Calcium-Aluminum rich inclusions (CAI)
and chondrule is known \citep{scott06}.
If the most of planetesimals are formed of chondrules,
dust aggregates as precursors of chondrules need to be retained 
in a disk during 1 Myr.
However, the gas friction makes the mm-sized dust 
in a laminar disk  to fall within 0.1Myr.
If the disk is turbulent, some of the dust may avoid falling
due to the turbulent diffusion.
Thus, even though the global turbulence is weak in the dead zone,
the turbulence due to the  instability described in this paper may 
play the role of avoiding planetesimal formation and 
floating  dust  in the disk.

The instability induced by the relative motion between gas and dust
should be studied more detailedly
because this instability have possible important roles 
in the dust evolution and the planetesimal formation
in the protoplanetary disk.

\acknowledgments
The calculations in this work were partly performed with computers at Astronomical Data Analysis Center, National Astronomical Observatory of Japan. 
This work was supported by Ministry of Education, Culture,
Sports, Science and Technology of Japan (MEXT) Grant-in-Aid 
for Scientific Research on Priority Areas, ``Development
of Extrasolar Planetary Science'' (MEXT-16077202).

%\clearpage 

%% Tables should be submitted one per page, so put a \clearpage before

%% Two options are available to the author for producing tables:  the
%% deluxetable environment provided by the AASTeX package or the LaTeX
%% table environment.  Use of deluxetable is preferred.
%%

%% Three table samples follow, two marked up in the deluxetable environment,
%% one marked up as a LaTeX table.

%% In this first example, note that the \tabletypesize{}
%% command has been used to reduce the font size of the table.
%% Note also that the \label command needs to be placed 

\clearpage

\begin{deluxetable}{ccccccccccc}
\tabletypesize{\scriptsize}
\tablecaption{Parameters and results \label{tbl-1}}
\tablewidth{0pt}
\tablehead{
\colhead{run} & \colhead{$\rho_d(0)/\rho_g$}   & \colhead{$h_d/z_d$}   &
\colhead{$\tau_f \Omega_K$} &
\colhead{$N_X \times N_z$ \tablenotemark{a}}
  & \colhead{$L_x/z_d$} & \colhead{$L_y / z_d$} &
\colhead{$\delta t \Omega_K$}   & \colhead{$\tau_D \Omega_K$\tablenotemark{b}}
 & \colhead{$k \eta r$} & \colhead{$\omega_I/\Omega_K$} 
}
\startdata
const  & 1.0 & 0.0  & 1  & 256 $\times$ 256 &  $\pi$ 
 & 1.0  & 5.e-4 & $10^3$ & 1.9 & 0.12 \\
r1t3  & 1.0 & 0.5  &  $10^{-3}$  & 256 $\times$ 256 & 2 $\pi$ 
 & 6.0  & 5.e-4 & $10^3$ &48.6 & 1.00 \\
r1t0  & 1.0 & 0.5  & 1  & 256 $\times$ 256 & 2 $\pi$   
& 8.0  & 2.e-4 & $10^3$ & 10.8 & 1.09 \\
 \enddata

%% Text for table notes should follow after the \enddata but before
%% the \end{deluxetable}. Make sure there is at least one \tablenotemark
%% in the table for each \tablenotetext.

\tablenotetext{a}{The number of Fourier components}
\tablenotetext{b}{$\tau_D \Omega_K = z_d^2 \Omega_K  / \nu_D $}
%\tablecomments{}

\end{deluxetable}

%% If you use the table environment, please indicate horizontal rules using
%% Do not put multiple tabular environments within a single table.
%% The optional \label should appear inside the \caption command.

\end{document}